\documentclass{article}
\usepackage{amssymb,amsmath}

\def\vvec{\underline{v}}
\def\uvec{\underline{u}}

\def\omvec{\underline{\omega}}
\def\del{\underline{\nabla}\,}
\def\delsq{\nabla^2}

\def\Psivec{\underline{\Psi}}

\def\ervec{\underline{e}_r}
\def\ethvec{\underline{e}_{\theta}}

\def\ns{Navier{\textendash}Stokes }

\usepackage{graphicx}

\begin{document}
\title{A simple resolution of Stokes' paradox?\footnote{This version April 2009, with a first consideration of iteration of the solution.}}
\author{William T. Shaw\thanks{Corresponding author: Department of Mathematics
King's College, The Strand,
London WC2R 2LS, England; E-mail: william.shaw@kcl.ac.uk}}

\maketitle
\begin{abstract}
This paper proposes a solution to Stokes' paradox for asymptotically uniform viscous flow around a cylinder.  The existence of a {\it global} stream function satisfying a perturbative form of the two-dimensional Navier{\textendash}Stokes equations for low Reynolds number is established. This stream function satisfies the appropriate boundary conditions on both the cylinder and at infinity, but nevertheless agrees with Stokes' original results at finite radius as the Reynolds number tends to zero. 
The Navier{\textendash}Stokes equations are satisfied to a power-log power of the Reynolds number. The drag on the cylinder is calculated from first principles and the free parameter of the approach can be chosen to give good agreement with data on drag. In this revised working paper we put our approach on a firmer mathematical basis using the Helmholtz-Laplace equation as a linear approximation to the Navier{\textendash}Stokes system. In so doing we demonstrate the instability of the original paradox. We also demonstrate the absence of a paradox of Stokes-Whitehead class, and give further theoretical constraints on the free parameters of the model.
\end{abstract}



\section{Introduction}
The difficulty in establishing a sensible global solution to the problem of low $R_e$ (Reynolds number) viscous flow around 
simple objects, where the flow is uniform at infinity, has fascinated applied mathematicians for just over 150 years. 
Stokes (1851) established that there was no solution to the two-dimensional, steady, incompressible, \ns equations for asymptotically uniform 
flow around a cylinder in the biharmonic limit. This situation is now routinely described in modern tutorial discussions.
See, for example, Chapter 7 of Acheson (1990) for an exercise on Stokes' paradox and a discussion of the corresponding situation 
for the sphere. The biharmonic equation can be solved in a neighbourhood of the cylinder (i.e. a circle) by any stream function of the form
\begin{equation}
\psi = C\sin\theta(r \log r - \frac{1}{2}r + \frac{1}{2r})
\end{equation}
but there is no choice of $C$ for which $\psi \sim r \sin\theta$ for large $r$. The analysis of this problem has lead to several classic papers (Oseen, 1910; Lamb, 1911) and its understanding through the use of matched asymptotic expansions (MAE) is one of the triumphs of perturbation theory. The reader is referred to Van Dyke (1964) for his classic survey of the work of Kaplun (1957), Proudman and Pearson (1957) and other key references. 

However, the MAE approach, despite its immense power and diversity of expanding applications, does not give a clean resolution of the original difficulty in that such methods rely essentially on computing and then matching solutions to the problem defined on two regions: close to the cylinder and far from the cylinder. The purpose of this paper is to address the problem of finding {\it global} solutions for the low Reynolds number limit, i.e. to resolve the original paradox. There are, of course, other approaches to the paradox. Recently, Villas Boas \cite{mario} has considered the problem from a three-dimensional perspective and points out that there is then no paradox. 

\section{Viscous incompressible flow in 2D}

A large class of fluids can be characterized by their density, $\rho$, a scalar field not presumed to be constant, and their dynamic viscosity $\mu$. The flow is characterized by a velocity vector field $\underline{v}$, and an associated scalar pressure field $p$. Conservation of mass is expressed by the continuity equation
\begin{equation}
\frac{\partial \rho}{\partial t} + \del.(\rho \vvec) = 0 
\end{equation}
and the conservation of momentum is expressed by the Navier-Stokes equations\footnote{Here  $\delsq$ acting on {\it vectors} should be understood as the ordinary Laplacian acting on Cartesian components.}
\begin{equation}
\rho(\frac{\partial \vvec}{\partial t} + \vvec.\del \vvec) = -\del p + \mu \delsq \vvec
\end{equation}
If the fluid is incompressible in the sense that $\rho$ is a constant in both time and space, we have the condition:
\begin{equation}
\del.\vvec = 0
\end{equation}
To analyze matters further, we introduce the vorticity vector
\begin{equation}
\omvec = \del \times \vvec
\end{equation}
In the following discussion we demand incompressibility but allow for non-zero vorticity. Using simple identities from vector calculus the Navier-Stokes equations may then be recast in the form
\begin{equation}
\rho(\frac{\partial \vvec}{\partial t} - \vvec \times \omvec) + \del (p + \frac{1}{2} \rho \vvec^2
)= -\mu \del \times \omega .
\end{equation}
Taking the curl of this, we arrive at the vorticity equation
\begin{equation}
\frac{\partial \omvec}{\partial t} + \vvec.\del \omvec - \omvec.\del \vvec = \nu \delsq \omvec
\end{equation}
where the kinematic viscosity $\nu = \mu/\rho$.
\subsection{The stream function}
Since the velocity field is divergence-free, we may introduce a vector potential $\Psivec$ such that
\begin{equation}
\vvec = \del \times \Psivec
\end{equation}
and furthermore we may choose it so that it is divergence free:
\begin{equation}
\del.\Psivec = 0
\end{equation}
The vector potential  can be reduced to a single function when there is an appropriate symmetry. The resulting object is a stream function. For example, planar 2D flow is obtained by setting (and note that this automatically satisfies satisfies the divergence condition)
\begin{equation}
\Psivec = - \Psi(x, y,t) \underline{e}_z
\end{equation}
Next we note that under the assumption that $\Psi$ satisfies $\del.\Psivec = 0$ 
\begin{equation}
\omvec = -\delsq \Psivec =  \underline{e}_z \delsq \Psi
\end{equation}
and the vorticity equation becomes

\begin{equation}
\nu \nabla^4 \Psi - \frac{\partial\ }{\partial t}\nabla^2 \Psi=  \frac{\partial(\Psi, \delsq \Psi)}{\partial(X, Y)}
\end{equation}

For problems where it is possible to identify a natural length scale $L$ and a natural speed $U$, it is normal 
practice to perform a non-dimensionalization of the variables and introduce the Reynolds number $R_e = UL/\nu$. 
Then the Navier{\textendash}Stokes equation becomes (after rescaling the variables suitably to $\psi, x, y, \tau$):
\begin{equation}
\nabla^4 \psi - R_e  \frac{\partial\ }{\partial \tau}\nabla^2 \psi= R_e \frac{\partial(\psi, \delsq \psi)}{\partial(x, y)}
\end{equation}
Throughout this paper we shall work in units in which the {\it radius} $a$ of the cylinder is taken to be unity. In such units the Reynolds number
is here based on the radius\footnote{it is also common to base $R_e$ on the diameter.} and is given by $R_e = U/\nu$. The two 
stream functions are related by $\Psi = U \psi$ and furthermore $X=x, Y=y$ etc.

The old historical approach to the limiting case when $R_e \rightarrow 0$ and the flow is time-independent is to take the view that the non-linearities may then be ignored (provided the non-linear term is well behaved) and the time-independent Navier{\textendash}Stokes equations reduce to 
\begin{equation}
\nabla^4 \psi = 0
\end{equation}
which is the biharmonic limit, also known as Stokes flow. It is now  well known (see for example, Chapter 8 of Van Dyke (1964)) that the neglect of the non-linear terms can lead to inconsistencies, as is evidenced by the lack of any solution for asymptotically uniform two-dimensional flow past a cylinder. 

Here we introduce the scalar vorticity function $\omega = \nabla^2 \psi$ and write the Navier{\textendash}Stokes as the pair
\begin{equation}
\begin{split}
\omega & = \nabla^2 \psi\\
\nabla^2 \omega - R_e  \frac{\partial \omega}{\partial \tau}&= R_e \frac{\partial(\psi, \omega)}{\partial(x, y)}
\end{split}
\end{equation}
If we had exponential growth in vorticity, $\omega = e^{\mu \tau}\hat{\omega}(x,y)$, then the linearized form would be
\begin{equation}
\begin{split}
 e^{\mu \tau}\hat{\omega}(x,y) & = \nabla^2 \psi\\
\nabla^2\hat{\omega} - R_e \mu \hat{\omega}&= 0\end{split}
\end{equation}
or in terms of a single condition:
\begin{equation}
\nabla^4 \psi - R_e \mu \nabla^2 \psi=0\ , 
\end{equation}
which is the Laplace-Helmholtz equation.

\subsection{The instability of Stokes' paradox}
Let us focus temporarily on the linearized time-dependent case based on the Laplace-Helmholtz equation. We let $\varepsilon^2 = R_e \mu$. It is an elementary exercise (and we shall give equivalent details in a different context later) to establish that for all $\varepsilon >0$, the Laplace-Helmholtz equation
\begin{equation}
\nabla^4 \psi - \varepsilon^2 \nabla^2 \psi=0
\end{equation}
has a solution of the form
\begin{equation}
\psi = e^{\mu \tau}\left[r
-\biggl(1+
\frac{2
   K_1(\varepsilon )}{\varepsilon  K_0(\varepsilon
   )}
\biggr) 
\frac{1}{r}
+\frac{2 K_1(r \varepsilon )}{\varepsilon 
 K_0(\varepsilon )}
 \right] \sin (\theta ) \label{tdsf}
\end{equation}
that satisfies
\begin{equation}
\begin{split}
\psi = 0 = \frac{\partial \psi}{\partial r} \ {\rm if }\  r= 1\\
\psi \sim  e^{\mu \tau} r \sin(\theta) \  {\rm as}\  r \rightarrow \infty
\end{split}
\end{equation}
In other words, no matter how small the value of $\mu$, the paradox does not exist. 

To summarize: If we have some time-dependence in the simplified form of exponential growth in vorticity the linearization is precisely the Laplace-Helmholtz problem rather than the biharmonic problem. Addition of this type of extra variability, no matter how small, shows that the paradox evaporates, i.e. the existence of the paradox is unstable. Villas Boas \cite{mario}, who has considered the problem from a three-dimensional perspective, also demonstrates that the paradox evaporates when extra {\it spatial} variability is incorporated.

In subsequent analysis we will use the Laplace-Helmholtz model as a approximation for the time-{\it independent} 2D case, but where the modification is regarded as a simple way as approximating the combined effect of the non-linearities via a linear term. The question is as to whether we can do this in a sensible and self-consistent manner. We shall look at this in a variety of ways, starting with a rather {\it ad hoc} approach.

\section{Reorganizing the low Reynolds number\\ \ns equations}
The production of a global solution to the \ns equations requires a slightly unusual approach.
We take the time-independent equation
\begin{equation}
\nabla^4 \psi = R_e \frac{\partial(\psi, \delsq \psi)}{\partial(x, y)} \label{oldfour}
\end{equation}
as our starting point and note some obvious facts. First, if we have a $\psi_0$ satisfying 
Laplace's equation, then it satisfies equation Eq.~(\ref{oldfour})  identically. 
Second, in the high Reynolds number limit, although this is a singular limit, we note that the Jacobian term should then 
vanish identically, so that 
any solution of
\begin{equation}
\delsq \psi = f(\psi)
\end{equation}
satisfies  Eq.~(\ref{oldfour})  when $R_e \rightarrow \infty$. This will be true of course in the linear case when 
$f(\psi) \propto \psi$ and this last equation is of Helmholtz type. Now let $\lambda$ be a (possibly complex) number of order 1 and $\beta$ a parameter to be determined. In particular, if we consider solutions, $\psi_1$, of a parametrized Helmholtz equation in the form
\begin{equation}
\delsq \psi_1 = \lambda R_e^{2\beta} \psi_1
\end{equation}
then the Jacobian will vanish and we note that
\begin{equation}
\nabla^4 \psi_1 - R_e \frac{\partial(\psi_1, \delsq \psi_1)}{\partial(x, y)} \equiv \lambda^2 R_e^{4\beta} \psi_1
\end{equation}
Next suppose that we consider a more general stream function $\psi$ as 
\begin{equation}
\psi = \psi_0 + \psi_1
\end{equation}
Then a trivial calculation tells us that
\begin{equation}
\nabla^4 \psi - R_e \frac{\partial(\psi, \delsq \psi)}{\partial(x, y)} = \lambda^2 R_e^{4\beta} \psi_1 - \lambda R_e^{2\beta+1} \frac{\partial(\psi_0, \psi_1)}{\partial(x, y)}   \label{expandedns}
\end{equation}
Our approach is therefore to combine solutions for potential flow, $\psi_0$, with a solution $\psi_1$ of the Helmholtz equation Eq. (23) that also satisfies the fourth order equation given by Eq. (24) without the identically zero non-linear term:
\begin{equation}
\nabla^4 \psi_1 = \lambda^2 R_e^{4\beta} \psi_1
\end{equation}
When $R_e \rightarrow 0$ this PDE of course approaches the biharmonic equation provided $\beta >0$. Then Eq.~(\ref{expandedns}) offers the possibility that we can solve a PDE of the form:
\begin{equation}
\nabla^4 \psi - R_e \frac{\partial(\psi, \delsq \psi)}{\partial(x, y)} = O(R_e^k) f(x,y)
\end{equation}
where $k$ and $f$ can be calculated and analyzed. The introduction of the term of order $R_e^{4\beta}$ in the right side of Eq.~(\ref{expandedns}) may seem like an artificial device, but given that (a) we are seeking solutions of the problem as $R_e \rightarrow 0$; (b) we are not modifying higher derivatives in the \ns equations; (c) we have a valid perturbation equation in Eq. (28), we consider that it is valid to proceed with this approach.

\subsection{Other justifications}
The author recognizes that the argument in the previous sub-section is somewhat {\it ad hoc}, and initially gives us no idea how to justify the choice of $\lambda$ or $\beta$, or, as we shall introduce, the relevant composite parameter $\epsilon$. However, there is another other route to justifying this approach. Let $\omega = \nabla^2 \psi$. Then the full 2D time-independent incompressible Navier--Stokes equation is precisely
\begin{equation}
\nabla^2  \omega- R_e \frac{\partial(\psi,  \omega)}{\partial(x,y)}=0\ .
\end{equation}
We already know from the existence of the Stokes paradox that simple linearization by setting $R_e=0$ is hopeless. We might therefore consider starting from other linearizations, for example, the Helmholtz linearization
\begin{equation}
\nabla^2 \omega - \epsilon^2 \omega=0\ ,
\end{equation}
where $\epsilon$ has to be determined either from experiment or from deeper theoretical considerations. The linear starting point for analysis is then not the biharmonic system but the Helmholtz-Laplace equation:
\begin{equation}
\nabla^4 \psi - \epsilon^2 \nabla^2\psi=0\ .
\end{equation}
In principle then we can consider justifying the choice of $\epsilon$ by matching $ \epsilon^2 \omega$ with $R_e \frac{\partial(\psi, \omega)}{\partial(x,y)}$ as closely as possible under a suitable norm. We will prove in the next section that {\it for all $\epsilon>0$ the Helmholtz-Laplace equation admits a solution satisfying both the boundary conditions on the circle and the velocity condition at infinity}, obtained by considering elementary solutions that are linear combinations of solutions of the Laplace equation and solutions of the Helmholtz equation.

Note that this approach and our {\it ad hoc} scheme are linked by the relationship
\begin{equation}
\epsilon^2 = \lambda R_e^{2\beta}
\end{equation}
The determination of $\lambda$ and $\beta$ are interesting challenges. We shall see in Section 8 that there are good theoretical grounds for setting $\beta=1$. For now we will leave both parameters general.

\section{The case of asymptotically uniform flow past a circle}
We now consider the well-trodden route to the analysis of a stream function associated with a flow that is uniform at infinity and satisfies appropriate boundary conditions on $r=1$. Using polar coordinates, we therefore want
\begin{equation}
\psi(r,\theta) \sim r \sin\theta  \ \ {\rm as}\ \  r \rightarrow \infty
\end{equation}
and
\begin{equation}
\psi = 0 = \frac{\partial \psi}{\partial r}\ \  {\rm on}\ \  r=1 \label{circlebc}
\end{equation}
We build the solution for $\psi_0$ and $\psi_1$ as follows, under the assumption that $\psi_0$ satisfies the Laplace equation and $\psi_1$ the Helmholtz conditions. Any sum of the two will satisfy the Laplace-Helmholtz equation. Given that the angular behaviour at infinity is $\propto \sin\theta$ we make the standard assumption and seek solutions
\begin{equation}
\psi_i =  g_i(r) \sin\theta\ \ {\rm for}\ \ i=0,1
\end{equation} 
The potential flow part (as usual) will be taken to be
\begin{equation}
g_0 = 1 + B/r
\end{equation}
for some constant $B$. Our analysis will differ from Stokes' classic (Stokes, 1851) treatment in that $g_1$ does not satisfy the separated biharmonic equation. Instead we use the Helmholtz equation, which in separated form is just
\begin{equation}
g_1''(r) +\frac{1}{r}g_1'(r)-\left(\epsilon^2+\frac{1}{r^2}\right)
   g_1(r)=0
\end{equation}
where $\epsilon = \sqrt{\lambda}R_e^{\beta}$. The general solution to this ODE is given by
\begin{equation}
g_1(r) = \alpha K_1(\epsilon r) + \beta I_1(\epsilon r)
\end{equation}
where $I_1, K_1$ are standard ``modified'' Bessel functions. It is now quite clear that we can construct $g_1$ so as the preserve the boundary condition at infinity by setting $\beta = 0$. The $K_1$ function decays exponentially if $\epsilon>0$. We are left with two arbitrary constants that can be determined by satisfying the boundary conditions on the circle given by Eq.~(\ref{circlebc}). This is a matter of elementary algebra using some standard Bessel function identities. The final result for the total stream function can be simplified to:
\begin{equation}
\psi = \left[r
-\biggl(1+
\frac{2
   K_1(\epsilon )}{\epsilon  K_0(\epsilon
   )}
\biggr) 
\frac{1}{r}
+\frac{2 K_1(r \epsilon )}{\epsilon 
 K_0(\epsilon )}
 \right] \sin (\theta ) \label{streamfn}
\end{equation}
We note some interesting facts about this expression. First, it has the {\it right} behaviour as $r \rightarrow \infty$ provided $\epsilon >0$, as the Bessel function of $r$ decays exponentially. Second, if we keep $r$ fixed and finite and let $\epsilon \rightarrow 0_+$ we obtain
\begin{equation}
\psi \sim \frac{\sin\theta}{\log(2/\epsilon) - \gamma}  \left(r \log r - \frac{1}{2}r + \frac{1}{2r}\right) \label{nearcyl}
\end{equation}
and we recover a multiple of Stokes' (1851) solution satisfying the boundary conditions on the sphere (but not at infinity), and, furthermore, the multiple is now reminiscent of that arising from the methods of matched asymptotic expansions. The claim therefore is that it is Eq.~(\ref{streamfn}) that essentially resolves the paradox, as all the boundary conditions are satisfied, but the low Reynolds number limit for finite $r$ does not. The third observation is that in the neighbourhood of the surface of the sphere,
\begin{equation}
\psi \sim \sin\theta (r-1)^2 \frac{\epsilon K_1(\epsilon)}{K_0(\epsilon)}
\end{equation}
and that for small $\epsilon$ this is given by
\begin{equation}
\frac{(r-1)^2 \sin (\theta
   )}{-\log (\epsilon )+\log (2)-\gamma
   }+O\left(\epsilon ^2\right)
\end{equation}
We also note that the limit as $\epsilon \rightarrow \infty$  is just the potential flow limit:
\begin{equation}
\psi \sim (r - \frac{1}{r}) \sin \theta
\end{equation} 
{\it Mathematica} code for the evaluation of the stream function and velocity field is given in the Appendix.

\subsection{Choice of $\lambda$ and $\beta$}
At this stage we have no knowledge about how to fix the parameter $\epsilon =  \sqrt{\lambda}R_e^{\beta}$, and we now turn 
our attention to this issue and a more detailed analysis of Eq.~(\ref{expandedns}). We note first that our working assumption of a 
power dependence of $\epsilon$ on $R_e$ has not actually been necessary. 
All we need is that $\epsilon \rightarrow 0_+$ as $R_e \rightarrow 0$. We also note that the choice $\beta=1$ 
and $\lambda = 1/(4 e)$ recovers the result (Van Dyke, 1964) that the limiting stream function for fixed $r$ and 
small $R_e$ is, from Eq.~(\ref{nearcyl})
\begin{equation}
\psi \sim \frac{\sin\theta}{\log(4/R_e) + \frac{1}{2} - \gamma}  \left(r \log r - \frac{1}{2}r + \frac{1}{2r}\right)
\end{equation}
so that the low $R_e$  limit matches exactly the first term of the MAE result.\footnote{for analyses basing $R_e$ on the diameter we have $\log(8/R_e)$ etc. in the denominator.} Clearly other choices of $\lambda$ can be considered, as can other powers of $R_e$ or even a more general $\epsilon$ still. However, in Section 8 we will establish a theoretical basis for estimating these parameters, and argue that $\beta=1$.

\section{Analysis of the remainder}
We must now analyze the right side of Eq.~(\ref{expandedns}){\textendash}we call it $\Gamma${\textendash}and we do so without any assumption as the the form of $\epsilon$. 
We have, changing variables in the Jacobian to polar coordinates:
\begin{equation}
\Gamma = \epsilon^4 \psi^1 - R_e \frac{\epsilon^2}{r}\left[ \frac{\partial \psi_0}{\partial r}\frac{\partial \psi_1}{\partial \theta} -  \frac{\partial \psi_1}{\partial r}\frac{\partial \psi_0}{\partial \theta}  \right]
\end{equation}
Recall that we set $\psi_i = g_i(r) \sin\theta$. Having imposed the boundary conditions, we have
\begin{equation}
g_0 = r - \frac{1}{r}\left(1 + \frac{2 K_1(\epsilon)}{\epsilon K_0(\epsilon)}  \right), \ \ 
g_1 = \frac{2 K_1(\epsilon r)}{\epsilon K_0(\epsilon)} 
\end{equation}
It follows that
\begin{equation}
\Gamma = \epsilon^4 \sin\theta g_1 - R_e \epsilon^2 \sin\theta \cos\theta J(\epsilon, r) = \Gamma_{HA} +\Gamma_{NL}
\end{equation}
where $J$ is the reduced Jacobian:
\begin{equation}
J = \frac{1}{r}\left(g_1 \frac{\partial g_0}{\partial r} - g_0 \frac{\partial g_1}{\partial r}\right)
\end{equation}
and $\Gamma_{HA} =  \epsilon^4 \sin\theta g_1 $ denotes the `Helmholtz artifact' introduced by our approach, and $\Gamma_{NL}$ denotes the non-linear Jacobian term.
Some calculation with Bessel function identities leads to
\begin{equation}
J(\epsilon, r) = \frac{2 \left(K_0(\epsilon ) K_2(r \epsilon )-\frac{1}{r^2}K_0(r \epsilon ) K_2(\epsilon)\right)}{K_0(\epsilon )^2}
   \end{equation}
from which it is manifest that $J(\epsilon,1) = 0$ and the non-linear terms vanish on the cylinder boundary irrespective of the choice of $\epsilon$. Furthermore, the asymptotic behaviour of the Bessel functions tells us that provided $\epsilon>0$,
\begin{equation}
J(\epsilon, r) = O(\frac{1}{\sqrt{r}} \exp(-\epsilon r))
\end{equation}
as $r \rightarrow \infty$. We deduce that $\Gamma_{NL}$ is a bounded function for all $r$. What is its order as a function of $R_e$? A straightforward estimate may be given by looking in a neighbourhood of the cylinder. It is straightforward to establish that
\begin{equation}
\frac{\partial J}{\partial r}|_{r=1} = \left(\frac{2 K_1(\epsilon)}{K_0(\epsilon)}  \right)^2 \sim \frac{4}{\epsilon ^2 (-\log (\epsilon )+\log (2)-\gamma )^2}
\end{equation}
where the latter expression applies as $\epsilon \rightarrow 0$. We deduce that in the {\it immediate} neighbourhood of the cylinder
\begin{equation}
\Gamma_{NL} = O\left(\frac{R_e}{(\log\epsilon)^2}\right)
\end{equation}
which is $o(R_e)$ provided only that $\epsilon \rightarrow 0$ as $R_e \rightarrow 0$. Although $J$ grows to a maximum as $r$ is increased away from unity, before reaching a maximum and then decaying for large $r$, some numerical experiments with {\it Mathematica} confirm that the maximum of  $\epsilon^2 J$ slowly decreases as $\epsilon$ decreases to zero. So we can assert that $\Gamma_{NL}$ is well behaved and is $o(R_e)$. 
When we consider $\Gamma_{HA}$, it is easy to see that $g_1$ has a maximum on $r=1$, and that 
\begin{equation}
\Gamma_{HA}|_{r=1} = \epsilon^4 \sin\theta g_1(1) = O(\epsilon^2/\log\epsilon)
\end{equation}
as $\epsilon \rightarrow 0$. So this term also behaves. If we desire that the Helmholtz artifact tends to zero faster than the non-linear term (which is desirable for the credibility of our approach) it is then natural to specify $\epsilon$ in the form $\sqrt{\lambda}R_e^{\beta}$ and to demand that $\beta \ge 1$. In particular, the choice $\beta=1$ arranges that
\begin{equation}
\Gamma_{HA} =  O(R_e^2/\log R_e)\ ,\ \ \Gamma_{NL} = O(R_e/\log^2{R_e})
\end{equation}
and we have established that $\Gamma = o(R_e)$. It also decays exponentially at infinity. 

\section{The drag on the cylinder}
The calculation of the drag can be done by integrating the pointwise force on the cylinder over its surface. The pointwise force has two components. One involves the local rate of strain in the fluid, and the other
is the pressure force. The first requires a purely local calculation, but the second
requires an integration of the pressure equation from infinity to the cylinder. A question is how this second part can be carried out {\it without} any {\it global} representation of the flow field. For the {\it specific} case of low Reynolds number calculations with a certain symmetry there are ways around the problem that we shall exploit presently. To make these matters clear, we shall summarize a first-principles calculation of the force using the results for cylindrical polar coordinates for the rate of strain as given by Appendix A of Acheson (1990). The fluid velocity is given (in our units) by
\begin{equation}
\underline{u} = -\frac{U}{r}\frac{\partial \psi}{\partial \theta} \underline{e}_r + U \frac{\partial \psi}{\partial r} \underline{e}_{\theta}
\end{equation}
On the cylinder the rate-of-strain tensor has components $e_{rr} = 0 = e_{\theta \theta}$, and
\begin{equation}
2e_{r\theta} = U \frac{\partial^2 \psi}{\partial r^2} 
\end{equation}
The stress tensor $T_{ij} = -p \delta_{ij} + 2\mu e_{ij}$, so the force on the cylinder boundary is
\begin{equation}
\underline{t} = T.\underline{e}_r =  -p \underline{e}_r  + \mu U \frac{\partial^2 \psi}{\partial r^2} \underline{e}_{\theta}
\end{equation}
We project this into the $x$-direction and integrate over the circle to get the following formula for the drag $D_0$ (force per unit length on the cylinder):
\begin{equation}
D_0 = - \int_0^{2\pi}\left[p \cos\theta + \mu U \frac{\partial^2 \psi}{\partial r^2} \sin\theta \right]d\theta
\end{equation} 
For low $R_e$ we estimate the pressure using the momentum equation in the form
\begin{equation}
\del p = -\mu \del \wedge (\del \wedge \uvec )
\end{equation}
and a short calculation gives
\begin{equation}
\del p = \mu U \left\{-\frac{1}{r} \frac{\partial\ }{\partial \theta}(\delsq \psi)\ervec + \frac{\partial\ }{\partial r}(\delsq \psi)\ethvec \right\}
\end{equation}
We need to integrate this in from infinity to a general point on the circle $r=1$. In general this is awkward without
a global form of $\psi$. But if $\psi$ has the form $\psi = g(r) \sin\theta$ then we have
\begin{equation}
\delsq \psi = \sin\theta {\cal L}[g(r)]\  \ {\rm where}\ \ {\cal L} [g(r)] \equiv  \frac{1}{r}\frac{\partial\ }{\partial r}\left(r\frac{\partial g}{\partial r} \right)- 
\frac{g}{r^2}
\end{equation}
and the pressure equation becomes:
\begin{equation}
\del p = \mu U \left\{-\ervec \frac{1}{r} \cos\theta {\cal L}[g(r)] + \ethvec \sin\theta \frac{\partial\ }{\partial r}{\cal L}[g(r)]\right\}
\end{equation}
We do the integration from infinity along $\theta = \pi/2$ and then work around the circle $r=1$. Carrying this out we obtain
\begin{equation}
p(1,\theta) = p_{\infty} - \mu U \cos \theta \frac{\partial\ }{\partial r} {\cal L}[g(r)]|_{r=1}
\end{equation}
We now evaluate the total integral for the force per unit length to obtain
\begin{equation}
D_0 = \mu U \pi \left[\frac{\partial\ }{\partial r}{\cal L}[g(r)]-\frac{\partial^2 \!g(r)}{\partial r^2} \right]
\end{equation}
The drag coefficient for such a 2D problem is defined following Tritton (1959) as $C_D = |D_0/(\rho U^2)|$
and is now easily seen to be given by
\begin{equation}
C_D = \frac{\pi}{R_e} \left|\frac{\partial\ }{\partial r}{\cal L}[g(r)]-
\frac{\partial^2\!g(r)}{\partial r^2} \right|
\end{equation}
evaluated on $r=1$. Given the vanishing of $g(1)$ and $g'(1)$ this simplifies further to
\begin{equation}
C_D = \frac{\pi}{R_e} |g^{(3)}(1) |
\end{equation}
In the case given by the MAE approach the stream function is given in the neighbourhood of the cylinder by Eq. (1) and then $g$ and $C_D$ are given by
\begin{equation}
g(r) = C\left(r \log r -\frac{r}{2} + \frac{1}{2r}\right),\ \ C_D = \frac{4 \pi C}{R_e}
\end{equation}
In our new model $g$ as is given by the radial part of Eq.~(\ref{streamfn}), and this time, using some Bessel identities, we obtain
\begin{equation}
C_D = \frac{2 \pi}{R_e} \frac{\epsilon^2 K_2(\epsilon)}{K_0(\epsilon)}
\end{equation}
\begin{figure}[hbt]
\centerline{ \includegraphics[width=3.0in]{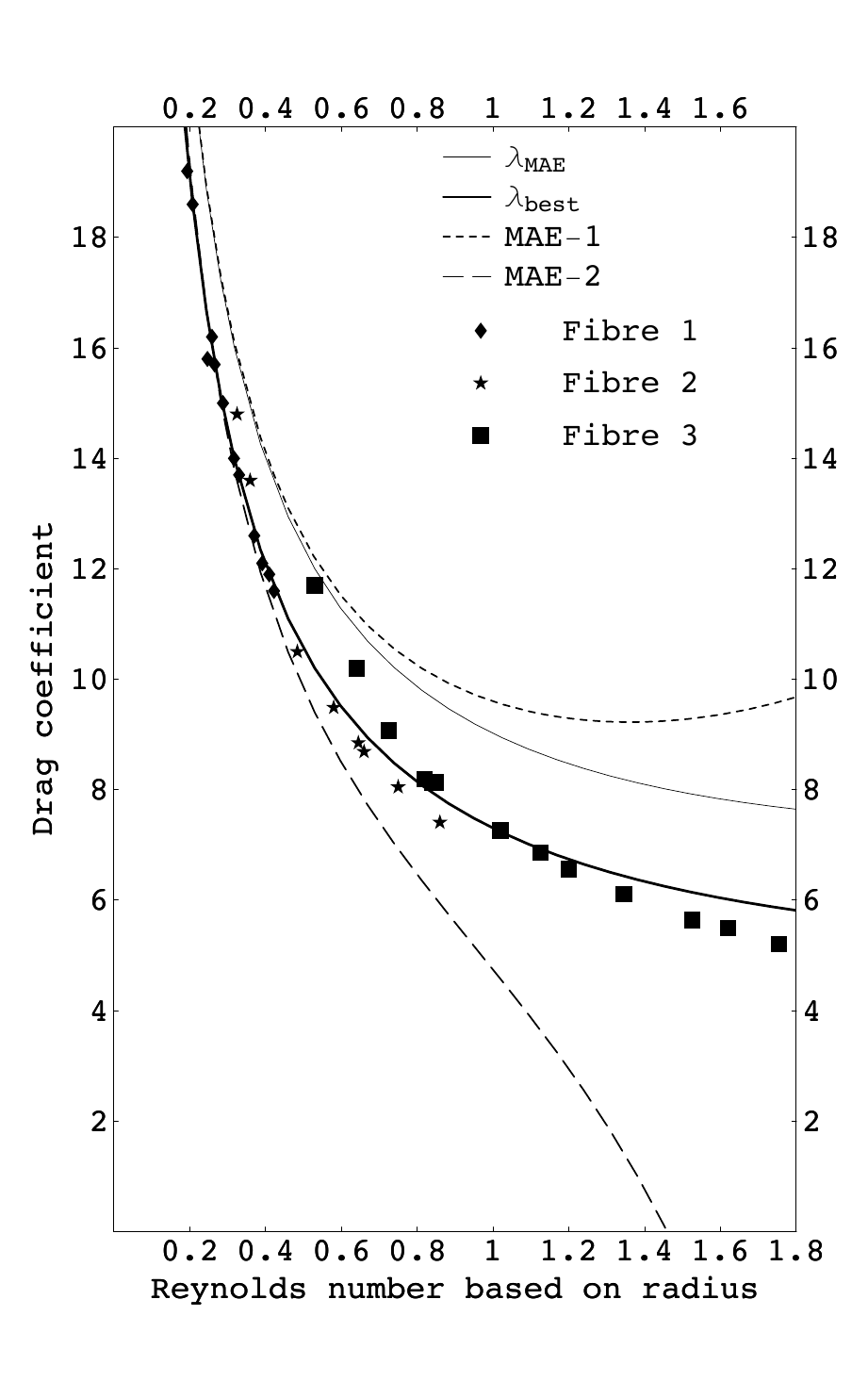}}
\caption{Comparison with Tritton (1959) data}
\end{figure}
We recall now the original assumption that $\epsilon = \sqrt{\lambda}R_e^{\beta}$. If we fix $\beta = 1$ but allow $\lambda$ to vary the drag coefficient of our model is then
\begin{equation}
C_D = 2 \pi \lambda R_e \frac{K_2(\sqrt{\lambda}R_e)}{K_0(\sqrt{\lambda}R_e)}
\end{equation}
We can plot the drag coefficient from our model, assuming that 
$\beta = 1$, with various choices for $\lambda$, and compare the results with those of Tritton (1959), which used an $R_e$ based on the diameter. Tritton's data has been converted by halving {\it his} Reynolds number. Also, in making a comparison with Fig. 8.5 of Van Dyke (1964), it appears that Van Dyke has plotted $C_D/(4 \pi)$, as otherwise it is not possible to reconcile that plot with Tritton's data. In Figure 1 we show the data for the first three fibres used by Tritton (those that consider the lowest $R_e$) and
\begin{itemize}
\item our model with $\lambda$ picked to match the MAE result, $\lambda = 1/(4 e) \sim 0.09197$;
\item our model with a least squares best fit $\lambda \sim 0.04452$;
\item the one term MAE result based on $ C = \Delta_1(R_e) = (\log(4/R_e) - \gamma + 1/2)^{-1}$;
\item the two term MAE result based on $ C = \Delta_1(R_e)- 0.8669 \Delta_1^3(R_e)$.
\end{itemize}
The choice of constants in the MAE approach is itself somewhat arbitrary. That made by Kaplun (1957), giving Eq.~(\ref{nearcyl}), is little more than convenience. These results give support to the new model and the agreement with experiment we get by taking $\lambda$ about one half that implied by the MAE approach is rather tantalizing. 

\newpage

\section{Iteration:  a Stokes-Whitehead anomaly?}
Another use for the type of low Reynolds numbers solution developed here is to provide a basis for {\it iteration} of the solution in powers of $R_e$. The presence of the paradox obstructing a global representation in the standard approach makes this impossible. Furthermore, we know from a corresponding analysis of the 3D spherical problem that even when the base solution makes sense, even the first iteration may fail: the {\it Stokes-Whitehead} paradox emerges. So it is of considerable interest to investigate the iteration of the system. In doing so we will keep $\epsilon$ general initially.

We summarize the solution developed thus far. With $\psi$ as given we have, combining equations (\ref{expandedns},45) and our definitions ,
\begin{equation}
\nabla^4 \psi - R_e \frac{\partial(\psi, \delsq \psi)}{\partial(x, y)} = \Gamma = \Gamma_{HA}+\Gamma_{NL}\ ,
\end{equation}
where
\begin{equation}
\Gamma_{HA} = \epsilon^4 \sin\theta g_1   =  \epsilon^4 \sin\theta \frac{2 K_1(\epsilon r)}{\epsilon K_0(\epsilon)} =\gamma_{HA}(\epsilon,r)  \sin\theta 
\end{equation}
and
\begin{equation}
\begin{split}
\Gamma_{NL} &=  - R_e \epsilon^2 \sin\theta \cos\theta J(\epsilon, r)\\ &=- R_e \epsilon^2 \sin\theta \cos\theta  \frac{2 \left(K_0(\epsilon ) K_2(r \epsilon )-\frac{1}{r^2}K_0(r \epsilon ) K_2(\epsilon)\right)}{K_0(\epsilon )^2}\\
&= \gamma_{NL}(\epsilon, r)\sin\theta \cos\theta 
\end{split}
\end{equation}
where these equations also serve to define the radial terms $\gamma_{XY}$. From now on we will work in terms of these radial functions, so it is useful just to write down their explicit forms:
\begin{equation}
\gamma_{HA}(\epsilon,r) = 2 \epsilon^3 \frac{K_1(\epsilon r)}{K_0(\epsilon)} 
\end{equation}
\begin{equation}
\gamma_{NL}(\epsilon,r) = -2 R_e \epsilon^2 \frac{ \left(K_0(\epsilon ) K_2(r \epsilon )-\frac{1}{r^2}K_0(r \epsilon ) K_2(\epsilon)\right)}{K_0(\epsilon )^2}
\end{equation}
\subsection{Development of the iteration}
We now attempt to refine the solution by writing the total solution as
\begin{equation}
\Psi = \psi + \psi_2
\end{equation}
Ideally we would like to arrange that
\begin{equation}
0 = \nabla^4 \Psi - R_e \frac{\partial(\Psi, \delsq \Psi)}{\partial(x, y)}
\end{equation}
The right side of this is given by the expansion
\begin{equation}
\nabla^4 \psi - R_e \frac{\partial(\psi, \delsq \psi)}{\partial(x, y)}+\nabla^4 \psi_2 - R_e \frac{\partial(\psi_2, \delsq \psi)}{\partial(x, y)}- R_e \frac{\partial(\psi, \delsq \psi_2)}{\partial(x, y)}
- R_e \frac{\partial(\psi_2, \delsq \psi_2)}{\partial(x, y)}
\end{equation}
and we write the remaining parts of the Navier-Stokes equation as
\begin{equation}
\begin{split}
0 = & \nabla^4 \psi_2  +\gamma_{NL}(\epsilon, r)\sin\theta \cos\theta +\gamma_{HA}(\epsilon,r)  \sin\theta \\
& - R_e \frac{\partial(\psi_2, \delsq \psi)}{\partial(x, y)}- R_e \frac{\partial(\psi, \delsq \psi_2)}{\partial(x, y)}
- R_e \frac{\partial(\psi_2, \delsq \psi_2)}{\partial(x, y)}
\end{split}
\end{equation}
We do not know how to solve this full non-linear system, and propose instead to treat a linearized form. The question now is whether to work with the raw system
\begin{equation}
0 =  \nabla^4 \psi_2  +\gamma_{NL}(\epsilon, r)\sin\theta \cos\theta +\gamma_{HA}(\epsilon,r)  \sin\theta \\
\end{equation}
or to propose a Helmholtz-damped variation along the lines of our original approach. We also need to understand which non-linear term to treat first. {\it Provided} $\epsilon$ is of the form $\epsilon = \sqrt{\lambda} R_e$, for example, we know that $\Gamma_{NL}$ is the lowest order in $R_e$: $\Gamma_{NL} = O(R_e^2/\log^2 R_e)$, so we will give priority to killing this lowest order correction. 

\subsection{Naive analysis of the raw forms}
To develop the solution we need to identify the Greens' s function for the biharmonic operator. In fact we do not need the full form as the angular structure of the right side is simple - we need only look for a pair of appropriate radial Green's functions satisfying appropriate boundary conditions on the cylinder and at infinity. To this end we write
\begin{equation}
\psi_2(r, \theta) = \psi_{3}(r) \sin\theta \cos\theta + \psi_4(r) \sin \theta
\end{equation}
and require that
\begin{equation}
\nabla^4 (\psi_{3}(r) \sin\theta \cos\theta) = - \gamma_{NL}(\epsilon, r)\sin\theta \cos\theta 
\end{equation}
and 
\begin{equation}
\nabla^4 (\psi_{4}(r) \sin\theta) = - \gamma_{HA}(\epsilon, r)\sin\theta 
\end{equation}
We also require that $\psi_i$ vanishes as $r \rightarrow \infty$ and both $\psi_i$ and its first derivative vanish on $r=1$.  To proceed further we recall the elementary form of the Laplacian in 2D cylindrical polar coordinates:
\begin{equation}
\nabla^2 f(r, \theta) = \frac{\partial^2 f}{\partial r^2} + \frac{1}{r} \frac{\partial f}{\partial r} + \frac{1}{r^2}  \frac{\partial^2 f}{\partial \theta^2}
\end{equation}
So if $f$ is of the form $f = h(r) \sin(k \theta)$ then 
\begin{equation}
\nabla^2 f(r, \theta) = \sin(k \theta) \biggl(\frac{d^2 h}{d r^2} + \frac{1}{r} \frac{dh}{dr} - \frac{k^2}{r^2} h\biggr)
\end{equation}
Define the $k$'the radial modal Laplacian operator as
\begin{equation}
{\cal L}_k = \frac{d^2 \ }{d r^2} + \frac{1}{r} \frac{d\ }{dr} - \frac{k^2}{r^2}
\end{equation}
We now have a pair of ODEs in the form
\begin{equation}
{\cal L}_2^2 \psi_3(r) =  - \gamma_{NL}(\epsilon, r)
\end{equation}
\begin{equation}
{\cal L}_1^2 \psi_4(r) =  - \gamma_{HA}(\epsilon, r)
\end{equation}
In a simpler problem we would now construct a pair of Green's functions $G_i(r,x)$, $i=1,2$,  with the properties that
\begin{equation}
{\cal L}_i^2 G_i(r,x) = \delta(r-x) 
\end{equation}
subject to, {\it if it is possible}, 
\begin{equation}
\lim_{r \rightarrow \infty}G_i(r,x) = 0
\end{equation}
\begin{equation}
G_i(1,x) = 0
\end{equation}
\begin{equation}
\frac{\partial\ }{\partial r}G_i(r,x)|_{r=1} = 0
\end{equation}
However, there are significant complications. The Helmholtz term we have thought of as an modelling artifact has first to be understood as an optimal linear approximation to the full non-linear system. The first steps on this are considered in Section 8. We will eventually have to manage the fact that no $G_1$ function exists satisfying these boundary conditions. However, it is now consistent to proceed to construct $G_2$, as this (a) exists, and (b) is needed for the management of the full non-linear system and its solution. 
\subsection{Integration of the inertia term}
We will now focus on the lowest order correction arising from the inertia term, for which $G_2$ is the required radial Green's function. The building blocks for this are two different solutions of the ODE
\begin{equation}
\biggl(\frac{d^2 \ }{d r^2} + \frac{1}{r} \frac{d\ }{dr} - \frac{2^2}{r^2}\biggr)^2 \phi(r) = 0\ .
\end{equation}
The solution to the homogeneous problem is of the form
\begin{equation}
\phi(r) = a r^4 + b r^2 + c + \frac{d}{r^2}
\end{equation}
To try to build a Green's function, we would write down a pair of solutions:
\begin{equation}
\begin{split}
\phi_L(r) &= a_L r^4 + b_L r^2 + c_L + \frac{d_L}{r^2}\ ,\ \ 1 \leq r \leq x\ ,\\
\phi_R(r) &= a_R r^4 + b_R r^2 + c_R + \frac{d_R}{r^2}\ ,\ \ x \leq r < \infty\ .
\end{split}
\end{equation}
The condition at infinity requires that we set $a_R = 0 = b_R$, so that
\begin{equation}
\phi_R(r) =  c_R + \frac{d_R}{r^2} ,\ \ x \leq r < \infty\ .
\end{equation}
The boundary conditions on the cylinder require that
\begin{equation}
\begin{split}
0 &= a_L+b_L+c_L+d_L \\
0 &= 4a_L + 2 b_L-2d_L
\end{split}
\end{equation}
So the inner function is given by
\begin{equation}
\phi_L(r)= \frac{2 a_L+b_L}{r^2}+a_L r^4+b_L r^2-2 b_L-3 a_L
   \end{equation}
 We now have four unknowns that must be obtained by imposing junction conditions to obtain the delta-function. The junction conditions are:
 \begin{equation}
 \begin{split}
\phi_R(x) &= \phi_L(x)\\
\phi'_R(x) &= \phi'_L(x)\\
\phi''_R(x) &= \phi''_L(x)\\
\phi_R^{(3)}(x) &= \phi_L^{(3)}(x)+1
\end{split}
 \end{equation}
These equations were solved and simplified using {\it Mathematica} V7 and the solutions are
\begin{equation}
\begin{split}
a_L &= --\frac{1}{48 x}\\
b_L &= \frac{1}{16}x\\
c_R &= \frac{\left(x^2-1\right)^2}{16x}\\
d_R &= -\frac{x^6-3 x^2+2}{48 x}
\end{split}
\end{equation}
Finally the two parts of the Green's function are (introducing the $x$-dependence explicitly)
\begin{equation}
\begin{split}
\phi_L(r,x) &=-\frac{r^4}{48 x}+\frac{r^2 x}{16}+\frac{\frac{x}{16}-\frac{1}{24x}}{r^2}-\frac{x}{8}+\frac{1} {16 x} \ ,\ \ 1 \leq r \leq x\ ,\\
\phi_R(r,x) &= -\frac{(x^6-3 x^2+2)}{48 r^2x}+\frac{\left(x^2-1\right)^2}{16x}\ ,\ \ x \leq r < \infty\ .
\end{split}
\end{equation}
The solution for $\psi_3$ is then given formally by
\begin{equation}
\psi_3(r) = -\int_1^r \gamma_{NL}(x) \phi_R(r,x)dx -\int_r^{\infty} \gamma_{NL}(x) \phi_L(r,x)dx
\end{equation}
This stream function is $O(1)$ as $r \rightarrow \infty$ so we claim that a secondary paradox of Stokes-Whitehead type has been avoided. To proceed further we need to better understand the full non-linear system. 

\section{The non-linear equations and constraints on $\epsilon$}
The task now at hand is to develop a proper theoretical basis for the estimation of the parameter $\epsilon$. To this end we must write down the full \ns equations. Converting to polar coordinates, the Navier--Stokes equations under consideration are
\begin{equation}
\nabla^4 \psi = \frac{R_e}{r} \frac{\partial(\psi, \delsq \psi)}{\partial(r, \theta)}
\end{equation}
Let us assume that $R_e$ is small enough that the flow remains symmetric about the horizontal axis. We may then write a Fourier decomposition
\begin{equation}
\psi = \sum_{m=1}^{\infty}\sin(m\theta) \phi_m(r) 
\end{equation}
Then with ${\cal L}_k$ as above, 
\begin{equation}
\nabla^2 \psi = \sum_{m=1}^{\infty}\sin(m\theta) {\cal L}_m \phi_m(r) 
\end{equation}
\begin{equation}
\nabla^4 \psi = \sum_{m=1}^{\infty}\sin(m\theta) {\cal L}_m^2 \phi_m(r) 
\end{equation}
\begin{equation}
\begin{split}
&\sum_{m=1}^{\infty}\sin(m\theta) {\cal L}_m^2 \phi_m(r) \\ &=  \frac{R_e}{r} \sum_{l=1}^{\infty} \sum_{n=1}^{\infty} \biggl[\sin(l\theta)n \cos(n\theta)\frac{\partial \phi_l}{\partial r}{\cal L}_n \phi_n - l \cos(l\theta) \sin(n\theta) \phi_l \frac{\partial \ }{\partial r}{\cal L}_n \phi_n  \biggr]\\
\\ &=  \frac{R_e}{2 r} \sum_{l=1}^{\infty} \sum_{n=1}^{\infty} \biggl[n[\sin((l-n)\theta)+\sin((l+n)\theta)]\frac{\partial \phi_l}{\partial r}{\cal L}_n \phi_n \\ &\ \ \ \ \ \ \ \ \ \ \ \ \ \ \ \ \ \ \ \ - l [\sin((l+n)\theta)-\sin((l-n)\theta)]\phi_l \frac{\partial \ }{\partial r}{\cal L}_n \phi_n  \biggr]
\end{split}
\end{equation}
Doing the Fourier analysis gives us
\begin{equation}
\begin{split}
&{\cal L}_m^2 \phi_m(r) \\ =  \frac{R_e}{2r}  \biggl[& (1-\delta_{m,1})\sum_{l=1}^{m-1}\biggl((m-l)\frac{\partial\phi_l}{\partial r}{\cal L}_{m-l}\phi_{m-l} - l \phi_l \frac{\partial \ }{\partial r}{\cal L}_{m-l}\phi_{m-l}\biggr)\\
&+\sum_{l=1}^{\infty}\biggl(l \frac{\partial \phi_{l+m}}{\partial r}{\cal L}_l \phi_l+(l+m)\phi_{l+m}\frac{\partial\ }{\partial r}{\cal L}_{l}\phi_{l}\biggr) \\
&-\sum_{l=1}^{\infty}\biggl((l+m) \frac{\partial \phi_l}{\partial r}{\cal L}_{l+m} \phi_{l+m}+l\phi_l\frac{\partial\ }{\partial r}{\cal L}_{l+m}\phi_{l+m}   \biggr)\biggr]
\end{split}
\end{equation}
Thus far no approximations have been made, other than to assume the flow remains symmetric about the $x$-axis. Now we approximate the model by considering only the contributions of the terms with $m=1,2$. We then have the simpler, but coupled and non-linear system
\begin{equation}
{\cal L}_2^2 \phi_2 = \frac{R_e}{2r}\biggl(\frac{\partial \phi_1}{\partial r}{\cal L}_1\phi_1 - \phi_1 \frac{\partial\ }{\partial r}{\cal L}_1 \phi_1  \biggr) \label{phitwonl}
\end{equation}
\begin{equation}
{\cal L}_1^2 \phi_1 = \frac{R_e}{2r}\biggl(\frac{\partial \phi_2}{\partial r}{\cal L}_1\phi_1 - \phi_1 \frac{\partial\ }{\partial r}{\cal L}_2 \phi_2 -2\frac{\partial \phi_1}{\partial r}{\cal L}_2\phi_2 +2 \phi_2 \frac{\partial\ }{\partial r}{\cal L}_1 \phi_1 \biggr)\label{phionenl}
\end{equation}
With our simple form of $\phi_1$ that is a solution of the Laplace-Helmholtz condition
\begin{equation}
{\cal L}_1^2 \phi_1 = \epsilon^2 {\cal L}_1 \phi_1\label{philin}
\end{equation}
we have already determined the solution of the Eq.~(\ref{phitwonl})-- $\phi_2$ is just $\psi_3$ as already given above. In order to generate maximal self-consistency of the linearization we must choose $\epsilon$ in such a way as to minimize the mismatch between the right sides of Eq.~(\ref{phionenl}) and Eq.~(\ref{philin})\footnote{This is of course one of many ways of proceeding, but our goal here is to provide a theoretical basis for the estimation of the so far free parameter $\epsilon$.}

Let us look at the overall scaling behaviour in terms of the Reynolds number. We let $\phi_2 = R_e \tilde{\phi}_2$, so that
\begin{equation}
{\cal L}_2^2 \tilde{\phi}_2 = \frac{1}{2r}\biggl(\frac{\partial \phi_1}{\partial r}{\cal L}_1\phi_1 - \phi_1 \frac{\partial\ }{\partial r}{\cal L}_1 \phi_1  \biggr)\end{equation}
The required matching is then of the form
\begin{equation}
\epsilon^2 {\cal L}_1 \phi_1 \sim {\cal L}_1^2 \phi_1 = \frac{R_e^2}{2r}\biggl(\frac{\partial \tilde{\phi}_2}{\partial r}{\cal L}_1\phi_1 - \phi_1 \frac{\partial\ }{\partial r}{\cal L}_2 \tilde{\phi}_2 -2\frac{\partial \phi_1}{\partial r}{\cal L}_2\tilde{\phi}_2 +2 \tilde{\phi}_2 \frac{\partial\ }{\partial r}{\cal L}_1 \phi_1 \biggr)\end{equation}
which strongly supports the scaling behaviour
\begin{equation}
\epsilon^2 = \lambda R_e^2
\end{equation}
as $R_e \rightarrow 0$. That is, we have a theoretical basis for setting $\beta=1$. A determination of $\lambda$ requires the introduction of a suitable rigorous criteria for minimizing the mismatch between the linear and non-linear forms, and this is under investigation.

Note that the {\it full} \ns equations, but limited to the first two angular modes, can certainly be written {\it without further approximation} as
\begin{equation}
{\cal L}_1^2 \phi_1 = R_e^2 \lambda(r) {\cal L}_1 \phi_1
\end{equation} 
for some unknown function $\lambda(r)$, and our method can now be properly understood as that of working with some ``average'' value of $\lambda(r)$, and noting that the resulting solution is free of a paradox.  Note also that writing the right side of this linearized system as a multiple of ${\cal L}_1 \phi_1$ is not as arbitrary as it might seem, for we know that the right side of the full non-linear system vanishes identically when ${\cal L}_1 \phi_1=0$. That is, this Laplace-Helmholtz model has a proper theoretical justification, rather than merely being the basis of a convenient interpolation between the boundary conditions on the cylinder and at infinity.

We could also consider generalizations where an improved {\it ansatz} for the form of $\lambda(r)$ is employed. The optimal average form and improved functional choices are under investigation. Only once this has been done would it make sense to consider further iteration. Some initial considerations suggest that as a function $\lambda(1)=0$ and that $\lambda(r)$ might be asymptotic to a constant independent of $R_e$ as $r \rightarrow \infty$, but further analysis is needed.
\section{Summary}
We have constructed a {\it global} stream function satisfying the two-dimensional viscous incompressible steady \ns equations in the limit $R_e \rightarrow 0$. The stream function satisfies the correct boundary conditions on a cylinder and infinity. The ``perturbation'' to the linearized \ns equations introduced to accomplish this is of higher order in $R_e$ than the inertia terms, and this modification is now properly understood as a linear representation of the full non-linear theory. The results suggest improved agreement with experimental data over those obtained by the MAE approach. Further work is needed on this approach, in particular on comparisons with newer data sets and determination of the remaining free parameter $\lambda$ from theoretical considerations. Although the approach of this paper initially started with rather {\it ad hoc} considerations, the method developed here is founded on a deeper consideration  based on the approximation of the non-linear \ns equations with an optimal linear approximation based on equations of Laplace-Helmholtz type, rather than the biharmonic equation. The need to properly treat the non-linearities has also been illustrated by the observation that the presence of a paradox is unstable with respect to small changes in the system. 

\subsection{Model summary}
The viscous and ``paradox-free'' stream function is given in non-iterated form by 
\begin{equation}
\psi = \left[r
-\biggl(1+
\frac{2
   K_1(\varepsilon )}{\varepsilon  K_0(\varepsilon
   )}
\biggr) 
\frac{1}{r}
+\frac{2 K_1(r \varepsilon )}{\varepsilon 
 K_0(\varepsilon )}
 \right] \sin (\theta ) 
 \end{equation}
 and the parameter $\epsilon = \sqrt{\lambda}R_e$.  The drag coefficient with Tritton's conventions is:
 \begin{equation}
C_D = \frac{2 \pi}{R_e} \frac{\epsilon^2 K_2(\epsilon)}{K_0(\epsilon)}
\end{equation}
The choice of $\lambda$ from theoretical considerations has yet to be determined, but experimental drag data suggests $\lambda \sim 0.04452$. The stream function satisfies the correct boundary conditions on the cylinder and at infinity for all $\epsilon>0$, and also reduces to that for potential flow as $\epsilon \rightarrow \infty$. {\it Mathematica} code for the stream function and velocity field are given in the Appendix. 

\subsection{Credit}
I am grateful to Bin Zhou for his comments on the earlier (2006) version of this paper.

\section*{Appendix: {\it Mathematica} code for the flow field}
The following code may be useful. First the stream function:
\begin{verbatim}
\[Psi][r_, \[Theta]_, \[Epsilon]_] := Sin[\[Theta]] (r - (1 + 
       2 BesselK[1, \[Epsilon]]/\[Epsilon]/BesselK[0, \[Epsilon]])/
     r + 2 BesselK[1, r \[Epsilon]]/\[Epsilon]/BesselK[0, \[Epsilon]])
\end{verbatim}
\begin{verbatim}
The streamlines are then easily visualized:
ContourPlot[\[Psi][Sqrt[x^2 + y^2], ArcTan[x, y], 0.01], {x, -10 , 
  10}, {y, -5, 5}, AspectRatio -> 1/2, Contours -> 50, 
 RegionFunction -> Function[{x, y}, x^2 + y^2 >= 1], 
 Epilog -> Circle[{0, 0}, 1]]
 \end{verbatim}
 
 \begin{figure}[hbt]
\centerline{ \includegraphics[width=3.5in]{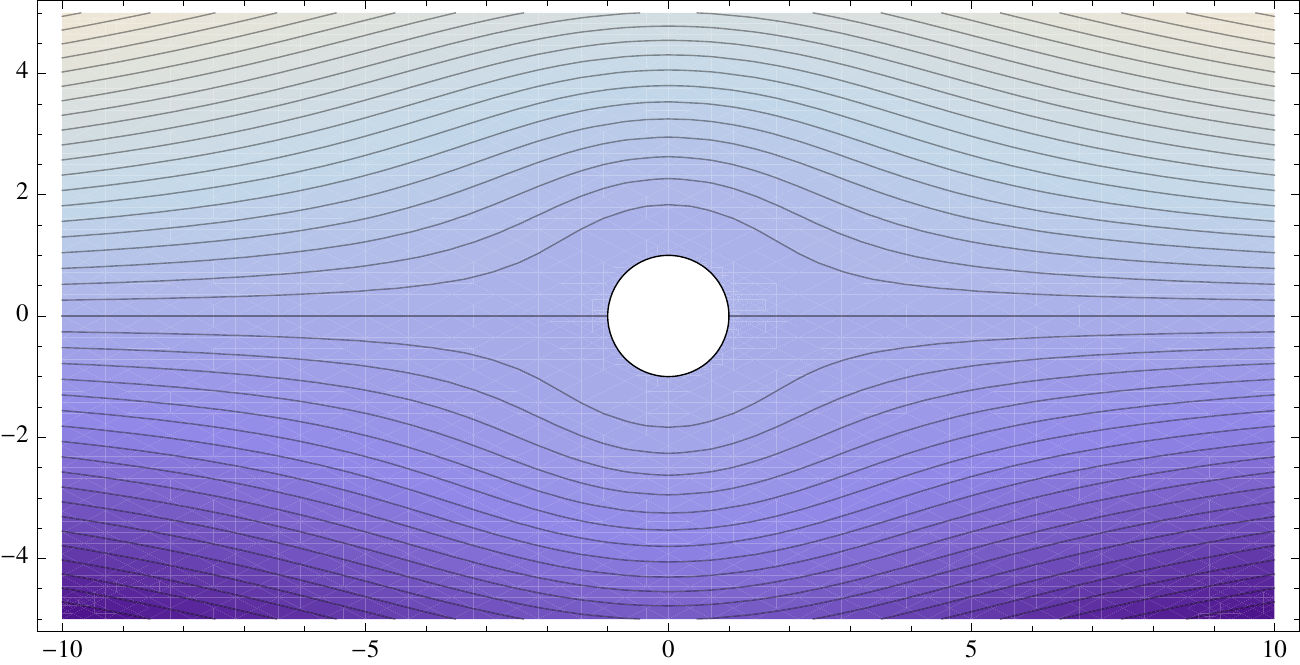}}
\caption{Streamlines}
\end{figure}
  The Cartesian components of the velocity field are given by
 \begin{verbatim}
 CartesianVelocity[x_, y_, U_, \[Epsilon]_] := 
 Module[{r = Sqrt[x^2 + y^2], \[Theta] = ArcTan[x, y], Ur, U\[Theta]},
   Ur = -((U*((-1 + r^2)*\[Epsilon]*BesselK[0, \[Epsilon]] - 
          2*BesselK[1, \[Epsilon]] + 2*r*BesselK[1, r*\[Epsilon]])*
        Cos[\[Theta]])/(r^2*\[Epsilon]*BesselK[0, \[Epsilon]]));
  U\[Theta] = 
   U*(1 + r^(-2) - (BesselK[0, 
          r*\[Epsilon]] - (2*
            BesselK[1, \[Epsilon]])/(r^2*\[Epsilon]) + 
         BesselK[2, r*\[Epsilon]])/BesselK[0, \[Epsilon]])*
    Sin[\[Theta]];
  Ur*{Cos[\[Theta]], Sin[\[Theta]]} + 
   U\[Theta]*{-Sin[\[Theta]], Cos[\[Theta]]}]
 \end{verbatim}
 The vector flow field is shown in Figure 3:
 \begin{verbatim}
 VectorPlot[
 If[x^2 + y^2 >= 1, 
  CartesianVelocity[x, y, 1, 0.1], {0, 0}], {x, -10, 10}, {y, -5, 5}, 
 RegionFunction -> Function[{x, y}, x^2 + y^2 >= 1], 
 AspectRatio -> 1/2 , VectorScale -> Small, 
 Epilog -> Circle[{0, 0}, 1]]
 \end{verbatim}
  \begin{figure}[hbt]
\centerline{ \includegraphics[width=3.5in]{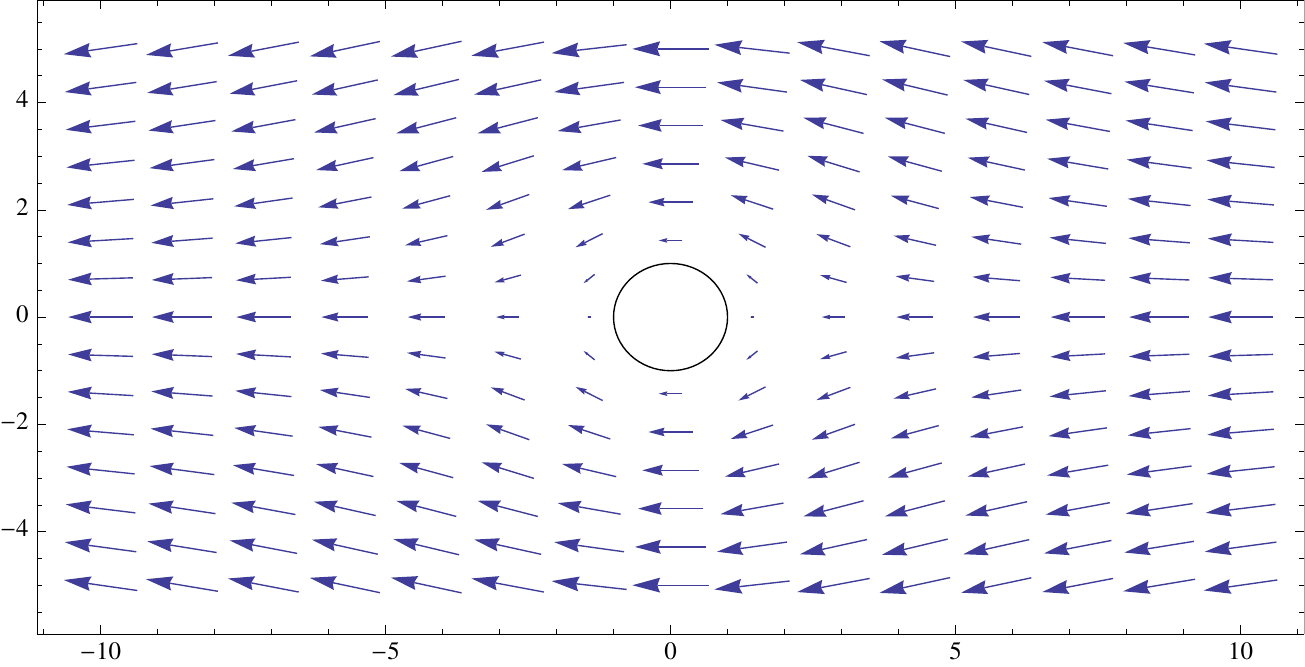}}
\caption{Vector field of flow}
\end{figure}
This illustrates the satisfaction of the boundary conditions on the cylinder. 
\end{document}